\documentclass[prd,aps,showpacs,secnumarabic,superscriptaddress]{revtex4}
\usepackage[dvips]{graphicx,color}
\def\al{\alpha}

\def\ga{\gamma}

\def\om{\omega}

\def\ti{\times}

\begin{document}
\title{Note on recent measurements of the $\psi(1S)\to\ga\,\eta_C(1S)$ \\ and $\psi(2S)\to\ga\,\eta_C(1S)$ branching ratios}
\author{Stanley F. Radford}
\affiliation{Department of Physics, The College at Brockport, State University of New York, \\ Brockport, New York 14420, USA} \email{sradford@brockport.edu}
\author{Wayne W. Repko}
\affiliation{Department of Physics and Astronomy, Michigan State University, East Lansing, Michigan 48824, USA} \email{repko@pa.msu.edu}
\date{\today}

\begin{abstract}
    Recently published measurements of the branching ratios ${\cal B}(\psi(1S)\to\ga\,\eta_C(1S))$ and ${\cal B}(\psi(2S)\to\ga\,\eta_C(1S))$ by the CLEO collaboration are examined in the context of a potential model that includes both relativistic and one-loop QCD corrections to the quark-antiquark interaction. The prediction for the width $\Gamma(\psi(1S)\to\ga\,\eta_C(1S))$ is in excellent agreement with the new data but the prediction for $\Gamma(\psi(2S)\to\ga\,\eta_C(1S))$ is too small. In an effort to understand this discrepancy, we derive an upper bound on $\Gamma(\psi(2S)\to\ga\,\eta_C(1S))$ and point out its experimental value saturates this bound.
\end{abstract}
\maketitle

\section{Introduction}

In a recent publication \cite{CLEO}, the CLEO collaboration reports new measurements of the branching ratios of the charmonium radiative decays $\psi(1S)\to\ga\,\eta_C(1S)$ and $\psi(2S)\to\ga\,\eta_C(1S)$. The new values are ${\cal B}(\psi(1S)\to\ga\,\eta_C(1S))=(1.98\pm 0.09\pm 0.30)\ti 10^{-2}$ and ${\cal B}(\psi(2S)\to\ga\,\eta_C(1S))=(4.32\pm 0.16\pm 0.60)\ti 10^{-3}$. These results imply radiative widths of  $\Gamma_{\rm exp}(\psi(1S)\to\ga\,\eta_C(1S))=1.85\pm 0.28$ and $\Gamma_{\rm exp}(\psi(2S)\to\ga\,\eta_C(1S))=1.41\pm 0.21$, both of which are larger than the current Particle Data Group values \cite{pdg}.

Magnetic dipole ($M1$) transitions of this type particularly sensitive to the details of the charmonium radial wave functions and, as such, these data provide an important check on model calculations of radiative transitions. There are numerous approaches to these calculations including lattice QCD \cite{LQCD}, heavy quark effective theory \cite{HQET}, inclusion of hadronic loop effects \cite{LZ} and potential models \cite{sr}. In this note, we describe the results of comparing the new data with the potential model calculations contained in Ref.\,\cite{sr}.

\section{M1 transitions}

In the dipole approximation, the width for the radiative transition $\psi(nS)\to\ga\,\eta_C(n'S)$ is given by
\begin{equation}\label{dip}
\Gamma(\psi(nS)\to\ga\,\eta_C(n'S))=\frac{4}{3}\frac{\al e_q^2}{m_c^2} \om^3|\langle n'00|n01\rangle|^2\frac{E_{\eta_C(n'S)}}{M_{\psi(nS)}}\,,
\end{equation}
where $\langle n'\ell s'|n\ell s\rangle$ denotes the radial integral
\begin{equation} \label{amps}
\langle n'\ell s'|n\ell s\rangle=\int_0^\infty dr r^2R_{n'\ell s'}(r)R_{n\ell s}(r)\,,
\end{equation}
where $\om$ is the photon energy and $E_{\eta_C(n'S)}$ is the energy of the recoiling $\eta_C$. Here $s$ denotes the initial spin $(s=0,1)$, $s'$ the final spin and $s'=s\pm 1$. In a model in which the radial wave functions used to compute the $M1$ matrix elements are obtained using a Hamiltonian that does not contain spin-dependent terms, the $\ell=0$ singlet and triplet radial wave functions corresponding to different radial excitations are themselves orthogonal. However, in general, $\ell=0$ singlet states $(s=0)$ and $\ell=0$ triplet states $(s=1)$ are orthogonal by virtue of their spin wave functions so there is no reason why the singlet radial functions $R_{n'00}(r)$ should be orthogonal to the triplet radial wave functions $R_{n01}(r)$ when $n'\neq n$. Given this, the radial wave function of the $\eta_C(1S)$ obtained in our non-perturbative treatment can be expanded in terms of the $\psi(nS)$ radial wave functions wave functions as
\begin{equation}
R_{100}(r)=\sum_{n=1}^\infty C_nR_{n01}(r)\,,
\end{equation}
with the $C_n$'s given by
\begin{equation}
C_n=\int_0^\infty dr r^2R_{n01}(r)\,R_{100}(r)\,.
\end{equation}
Hence, the amplitudes $\langle 100|n01\rangle$ are just the $C_n$'s and, since the $\eta_C(1S)$ radial wave function is normalized,
\begin{equation}\label{norm}
\sum_{n=1}^\infty |C_n|^2=1\,.
\end{equation}
Given a model that adequately describes $\Gamma(\psi(1S)\to\ga\,\eta_C(1S))$, the value of $|C_1|^2$ can be used to obtain a bound on  $\Gamma(\psi(2S)\to\ga\,\eta_C(1S))$ by noting that Eq.\,(\ref{norm}) can be written
\begin{equation} \label{bound}
1-|C_1|^2=\sum_{n=2}^\infty |C_n|^2\geq |C_2|^2\,.
\end{equation}

To see if the bound in Eq.\,(\ref{bound}) is particularly restrictive, we use the results of the potential model described in detail in Ref.\,\cite{sr}. Briefly, the model uses a Hamiltonian consisting of a short distance potential that includes all $v^2/c^2$ relativistic corrections and one-loop QCD corrections, a linear confining potential that is a mixture of scalar and vector contributions together with their $v^2/c^2$ corrections and a relativistic kinetic energy term. The charmonium spectrum is obtained using a variational technique that provides explicit forms for the radial wave functions. The amplitudes in Eq.\,(\ref{amps}) can then be evaluated and the predictions for the radiative widths calculated using Eq.\,(\ref{dip}).

The radiative widths calculated this way using the non-perturbative approach of Ref.\,\cite{sr} are
\begin{eqnarray}
\Gamma(\psi(1S)\to\ga\,\eta_C(1S)) &=& 1.84\; {\rm keV}\,, \label{1S}\\
\Gamma(\psi(2S)\to\ga\,\eta_C(1S)) &=& 0.44\; {\rm keV}\,. \label{2S}
\end{eqnarray}
From Eq.\,(\ref{1S}), the width of the $\psi(1S)\to\ga\,\eta_C(1S)$ transition is well described by the model, but Eq.\,(\ref{2S}) shows a discrepancy of about a factor of 3. Faced with this discrepancy, one might ask if it is possible to modify the $\psi(2S)\to\ga\,\eta_C(1S)$ amplitude enough to achieve agreement with experiment without violating probability conservation. The extent to which this can be done is controlled by Eq.\,(\ref{bound}). The calculated value of $|C_1|^2$ leading to the result in Eq.\,(\ref{1S}) is $|C_1|^2=0.9958$, so Eq.\,(\ref{bound}) becomes
\begin{equation} \label{C2}
|C_2|^2\leq 4.196\ti 10^{-3}\,,
\end{equation}
whereas the calculation leading to Eq.\,(\ref{2S}) gives $|C_2|^2=1.571\ti 10^{-3}$. Rescaling the result in Eq.\,(\ref{2S}) by the ratio then implies
\begin{equation} \label{gbnd}
\Gamma(\psi(2S)\to\ga\,\eta_C(1S))\leq 1.20\; {\rm keV}\,.
\end{equation}
Interestingly, the experimental value, $\Gamma_{\rm exp}(\psi(2S)\to\ga\,\eta_C(1S))=1.41\pm 0.21$, saturates this bound within errors. Before drawing any conclusions, it must be remembered that the bound is obtained using the dipole approximation, which amounts to replacing the complete $M1$ amplitude,
\begin{equation}\label{jamp}
{\cal A}(M1)=\int_0^\infty dr r^2R_{n'\ell s'}(r)R_{n\ell s}(r)j_0(\frac{\om r}{2})\,,
\end{equation}
with Eq.\,(\ref{amps}) by using $j_0(x)\stackrel{{x\to 0}}{\longrightarrow} 1$. This is justified as long as the correction from the next term in the expansion of $j_0(\frac{\om r}{2})$, $\om^2r^2/4!$, is small. In the present case, the wave functions used in the variational calculation fall off exponentially at large $r$ with a scale parameter $R=1\;{\rm Gev^{-1}}$. Using $R$ to estimate the size of the correction, we find $5.32\ti 10^{-4}$ for the $\psi(1S)\to\ga\,\eta_C(1S)$ transition and $1.69\ti 10^{-2}$ for $\psi(2S)\to\ga\,\eta_C(1S)$ transition. This indicates that deviations from the dipole approximation do not materially affect the validity of the bound on $\Gamma(\psi(2S)\to\ga\,\eta_C(1S)$ given in Eq.\,(\ref{gbnd}).

\section{Conclusion}

Within the confines of the dipole approximation, a bound of the type in Eq.\,(\ref{bound}) on the $M1$ transition amplitude from the first excited triplet state to the singlet ground state can always be obtained. Furthermore, the transition amplitudes in any sensible model calculation will satisfy this bound.

So, is there reason to be concerned that $\Gamma_{\rm exp}(\psi(2S)\to\ga\,\eta_C(1S))$ just barely satisfies the bound associated with a particular model that accurately predicts $\Gamma_{\rm exp}(\psi(1S)\to\ga\,\eta_C(1S))$? From the potential model point of view, the value of $\Gamma_{\rm exp}(\psi(2S)\to\ga\,\eta_C(1S))$ is uncomfortably large because all the ingredients needed to evaluate Eq.\,(\ref{dip}) are tightly constrained by the requirement that the variational calculation accurately reproduce the observed charmonium spectrum \cite{sr}. Once this is accomplished, the $|C_n|^2$'s are determined and those with $n>2$ are not zero. The challenge posed by the new data is one of seeing whether it is possible to refine the wave functions in such a way that $|C_2|^2$ can be increased within the constraints of probability conservation without sacrificing the quality of the overall fit to the charmonium spectrum.

\begin{acknowledgments}
This work was supported in part by the National Science Foundation under Grant PHY-0555544.
\end{acknowledgments}

\end{document}